\documentstyle[12pt]{article}
\def\fun#1#2{\lower3.6pt\vbox{\baselineskip0pt\lineskip.9pt
  \ialign{$\mathsurround=0pt#1\hfil##\hfil$\crcr#2\crcr\sim\crcr}}}
\newskip\humongous \humongous=0pt plus 1000pt minus 1000pt

\newif\ifdtup

\def\oldreffmt#1{\rlap{[#1]} \hbox to 2\parindent{}}

\def\figfmt#1{\rlap{Figure {#1}} \hbox to 1in{}}

\def\beq{\begin{equation}}
\def\eeq{\end{equation}}

\def\bq{\begin{quote}}
\def\eq{\end{quote}}

\relax

\newcommand{\be}{\begin{equation}}
\newcommand{\ee}{\end{equation}}
\addtocounter{equation}{-1}
\begin{document}
\baselineskip=18pt plus 1pt minus 1pt

\begin{titlepage}
\rightline{CPT-92/P.2679}
\rightline{April 1992}
\vskip 1.5truecm
\begin{center}
{\large{\bf Lie-algebraic approach to the theory of polynomial solutions.\\
I. Ordinary differential equations and finite-difference equations in one
 variable}}
\vskip 0.6truecm
(submitted to Comm.Math.Phys.)
\vskip 0.6truecm
 {\bf A.Turbiner}\footnote
{On leave of absence from: Institute for Theoretical and Experimental Physics,
Moscow 117259, Russia\\E-mail: TURBINER@CERNVM or TURBINER@VXCERN.CERN.CH}
\vskip 0.5cm
CPT, CNRS-Luminy, Marseille, F-13288, FRANCE
\\
\end{center}

\vskip 1.2cm
\begin{center}
{\large ABSTRACT}
\end{center}
\vskip 0.5 cm
\begin{quote}

A classification of ordinary differential equations
 and finite-difference equations in one variable having
polynomial solutions (the generalized Bochner
problem) is given. The method used is
based on the spectral problem for a polynomial element of
the universal enveloping algebra of $sl_2({\bf R})$ (for differential
equations) or $sl_2({\bf R})_q$ (for finite-difference equations)
in the "projectivized" representation possessing an  invariant subspace.
Connection to the recently-discovered quasi-exactly-solvable
problems is discussed.
\end{quote}

\vfill

\end{titlepage}
\newpage

 S. Bochner (1929) asked about a classification of differential equations
\be
 T \varphi  \  = \ \epsilon \varphi
\ee
where $T$ is a linear differential operator of $k$-th order in one real
variable $x \in {\bf R}$
and $\epsilon$ is the spectral parameter, having an infinite sequence of
{\it orthogonal} polynomial solutions (see \cite{little}).

{\bf Definition.} Let us give the name of the {\it generalized Bochner
 problem} to the problem of classification of the differential equations
 (0) having $(n+1)$ eigenfunctions in the form of a polynomial of the order
not higher than $n$.

In Ref.2 it  has been formulated a general method for generating
 eigenvalue problems for linear differential
 operators, linear matrix differential operators and
 linear finite-difference operators in one and several variables
possessing the polynomial solutions. The
method is based on considering the eigenvalue problem for the representation of
 a polynomial element
of the universal enveloping algebra of the Lie algebra in a
finite-dimensional, 'projectivized'
representation of this Lie algebra \cite{t1}. Below it is shown that
 this method provides both necessary and sufficient conditions for the
existance
of polynomial solutions of linear differential equations
and a certain class of finite-difference  equations in one variable.

{\bf 1. Ordinary differential equations.}

Consider the space of all polynomials of order $n$
\be
\label{e1}
{\cal P}_n \ = \ \langle 1, x, x^2, \dots , x^n \rangle ,
\ee
where $n$ is a non-negative integer and $x \in {\bf R}$. \par

{\bf Definition.} Let us name a linear differential operator of the $k$-th
order, $T_k$ {\bf quasi-exactly-solvable}, if it preserves the space ${\cal
P}_n$. Correspondingly, the operator $E_k$, which preserves
the infinite flag $ {\cal P}_0 \subset  {\cal P}_1 \subset {\cal P}_2
\subset \dots \subset {\cal P}_n \subset \dots $ of spaces of all
polynomials, is named {\bf exactly-solvable}.

{\bf LEMMA 1. } {\it (i) Suppose $n > (k-1)$.  Any quasi-exactly-solvable
operator of $k$-th order $T_k$, can be
represented by a $k$-th degree polynomial of the operators}

\label{e2}
\[ J^+ = x^2 \partial_x - n x,\  \]
\be
 J^0 = x \partial_x - {n \over 2} \  ,
\ee
\[ J^- = \partial_x \ ,  \]
\noindent
{\it (the operators (2) obey the $sl_2({\bf R})$ commutation relations}
\footnote{The representation (2) is one of the 'projectivized'
representations (see \cite{t1}).}
{\it). If $n \leq (k-1)$, the part of the quasi-exactly-solvable operator
$T_k$ containing
derivatives up to the order $n$ can be represented by a $n$-th
degree polynomial in the generators (2).

(ii) Inversely, any polynomial in (2) is quasi-exactly solvable.

(iii) Among quasi-exactly-solvable operators
there exist exactly-solvable operators $E_k \subset T_k$.}\par

{\it Comment 1.} If we define the universal enveloping algebra $U_g$ of a
Lie algebra $g$ as the algebra of all polynomials in generators, then
$T_k$ at $k < n+1$ is simply an element
 of the universal enveloping algebra $U_{sl_2({\bf R})}$ of the algebra
$sl_2({\bf R})$ taken in representation (2). If $k \geq n+1$, then $T_k$ is
represented as an element of $U_{sl_2({\bf R})}$ plus $B {d^{n+1}
\over dx^{n+1}}$, where $B$ is any linear differential operator of the order
not higher than $(k-n-1)$. \par

Now let us proceed to the proof of the Lemma 1.

{\bf Proof.}

Part I. One starts from the statement (i). Suppose $n>(k-1)$.

{\it Step 1.} It is easy to show that the coefficient functions standing
before derivatives in any
quasi-exactly-solvable operator $T_k$ should be polynomials.

{\it Step 2.} Without loss of generality, any operator $T_k$ can be represented
as a sum of homogeneous operators:
\be
\label{e3}
T_k = \sum_{i=0}^k a_{k,i}^{(n_i)} (x) \partial _x^{i}
\ee
where $n_{i}$ indicates the degree of polynomial $ a_{k,i}^{(n_{i})} (x)$.
Suppose, the maximal degree of homogenuity is $M=n_{i}-i >0$. Now let us
rewrite $T_k$ as sum of operators of fixed degree of homogenuity $m$:
\be
\label{e4}
T_k = \sum_{m=-k}^M T_k^{(m)} \ , \ T_k^{(m)} \equiv \sum_{i=0}^k
A_{k,i}^{(m)} x^{m+i} \partial_x^{i}
\ee
(if $m+i<0$, the corresponding $A_{k,i}^{(m)}=0$). Evidently, if $m \leq 0$,
 then $T_k^{(m)} : {\cal P}_n \mapsto {\cal P}_n$. Consider the operators
$T_k^{(m)}$ at $m>0$. Their action on monomials is
\be
\label{e5}
T_k^{(m)} : x^{\ell} \mapsto x^{\ell + m}
\ee
In order to preserve the space (1), the condition
\be
\label{e6}
\ell + m \leq n \ , \ \ell = 0,1,2, \dots n
\ee
should be fulfilled.
Generically, $T_k^{(m)}$ is characterized by $(k+1)$ coefficients. The
condition (4) leads to the system of $m$ linear homogeneous equations for
$(k+1)$ unknown coefficients $A_{k,i}^{(m)}$. Hence
\be
\label{e7}
m \leq k
\ee
and the maximal degree of homogenuity of a quasi-exactly-solvable operator
$T_k$ is equal to k.

{\it Step 3.} We will prove (i) inductively.

{\it (1).} Take $T_0$. Evidently, $T_0 = const$ or, in other words,
a zero-order polynomial in generators (2).

{\it (2).} Take $T_1$. As it follows from (4),(7)
\be
\label{e8}
T_1=\sum_{m=-1}^1 T_1^{(m)} \equiv (A_{1,0}^{(1)} x +  A_{1,1}^{(1)} x^2
\partial_x ) + (A_{1,0}^{(0)} +  A_{1,1}^{(0)} x \partial _x ) +
( A_{1,1}^{(-1)} \partial _x ) \ .
\ee
We only need to consider the operator $T_1^{(1)}$, since the others preserve
 (1) automatically. In this case, the above-mentioned system of linear
equations contain only one equation:
\be
\label{e9}
A_{1,0}^{(1)} + n A_{1,1}^{(1)} \  = \ 0 \ ,
\ee
which implies immediately, that
\be
\label{e10}
T_1^{(1)}= A_{1,1}^{(1)} J^+ \ .
\ee
Finally,
\be
\label{e11}
T_1^{(1)}\ =\ A_{1,1}^{(1)} J^+\ +\ A_{1,1}^{(0)} J^0\ +\  A_{1,1}^{(-1)}
J^-\ + \ A_{1,0}^{(0)}\ +\  {n \over 2} A_{1,1}^{(0)}
\ee

{\it (3).} Now let us assume that a quasi-exactly-solvable operator $T_{k-1}$
is represented through generators (2).

 Take $T_k$. In what follows from
{\it Step 2}, the maximal degree of homogenuity of $T_k$ is equal to $k$.
First consider $T_k^{(k)}$. The corresponding system of linear equations on
the coefficients $A_{k,i}^{(k)}, i=0,1,2, \dots k$ contains $k$ equations
and hence has a non-trivial solution.
This solution can be expressed through one coefficient, e.g.
$A_{k,k}^{(k)}$. It is easy to check that obtained coefficients correspond
precisely to $(J^+)^k$ (cf.(9),(10)) and hence
\be
\label{e12}
T_k ^{(k)}=A_{k,k}^{(k)} (J^+)^k
\ee
It is obvious, that (12) preserves the space (1). The remainder of $T_k$ is
simply the $k$-th order differential operator characterizing by degrees of
homogenuity $-k,-k+1,\dots, 0,1, \dots,k-1$. Now let us consider in the
remainder the terms containing
the derivatives of the $k$-order. Apparently, they can be rewritten in the
following forms:
 \[x^{2k-1} \partial^k = (J^+)^{k-1} J^0 + \alpha_1 x^{2k-2} \partial^{k-1}+
\dots \]
 \[x^{2k-2} \partial^k = (J^+)^{k-1} J^- + \alpha_2 x^{2k-3} \partial^{k-1}+
\dots \]
 \[x^{2k-3} \partial^k = (J^+)^{k-2} J^0 J^- + \alpha_3 x^{2k-4}
\partial^{k-1}+ \dots \]
\[ \vdots \]
\be
\label{e13}
\partial^k = (J^-)^k
\ee
(dots in (13) imply terms with lower order derivatives). The
representation (13) is unambiguous, since there are only $(2k+1)$ different
monomials of degree $k$ in the generators $J^{\alpha}$ (if we take into
account the commuation relations and the elements of the ideal generated by the
Casimir operator). Substituting (12),(13) into $T_k$ in the form (3), one gets
\be
\label{e14}
T_k = \sum_{k_+,k_0,k_\geq 0} ^{k_++k_0+k_-=k} P_{k_+,k_0,k_-}
(J^+,J^0,J^-) \ + \ T_{k-1} \ .
\ee
The operator $T_{k-1}$ can be represented through (2) by assumption.
This ends the proof of the first part of (i).

Part II. Suppose $n \leq k-1$. It is clear that the part of $T_k$, containing
derivatives of the orders \ $(n+1),(n+2),\dots, k$ \ annihilates the space (1)
and these derivatives
can keep any functions as coefficient functions. For the remainder of $T_k$,
containing derivatives up to the $n$-th order, the Part I of the proof  holds.
This concludes the proof of part (i) of the Lemma.

Part III. The part (ii) of the Lemma is evident. The part (iii) of the Lemma
is also evident from the proof of part (i) of the Lemma, since a
quasi-exactly-solvable operator, containing no
operators of positive homogenuity, becomes exactly-solvable. $\Box$

\noindent
{\it Comment 2.} If the space (1) is considered not in general position
\footnote{\ 'not in general position' means that, it is allowed to have some
correlations between coefficients of original polynomial and an
operator $T_k$. Otherwords, not any polynomial of degree not higher than $n$
is mapped to ${\cal P}_n$.}, there exist linear differential operators others
than quasi-exactly-solvable ones, which map a certain polynomial
of a degree $n$ to some other polynomial of the same degree.


Since $sl_2({\bf R})$ is a graded algebra, let us introduce the grading
of generators (2):
\be
\label{e15}
deg (J^+) = +1 \ , \ deg (J^0) = 0 \ , \ deg (J^-) = -1 ,
\ee
 hence
\be
\label{e16}
deg [(J^+)^{n_+} (J^0)^{n_0}(J^-)^{n_-}] \  = \ n_+ - n_- .
\ee
The grading allows to classify the operators $T_k$ in Lie-algebraic sense.\par

{\bf LEMMA 2. } {\it A quasi-exactly-solvable operator $T_k \subset
U_{sl_2({\bf R})}$ has no terms of positive grading, iff it is an
exactly-solvable operator.} \par

\noindent
{\it Comment 3.} It is easy to see that the grading is nothing but the
homogenuity, which has been introduced in the proof of the Lemma 1 and
 the statement of this Lemma becomes obvious.

{\bf Definition.} Let us call the operator $T(x)$ symmetric, if one can
introduce the scalar product
\[ (f,g)_{\rho} = \int_R f(x) g(x) \rho (x) dx \]
for positive $\rho > 0, \rho (x) \in S(R)$, such that
\[ (Tf,g)_{\rho} = (f,Tg)_{\rho} \]
where the Schwatz space is defined as
\[ S(R)=\{ \rho : R \rightarrow R, \mid
\rho^{(k)} (x) \mid \leq C_{kn} (1+ \mid x \mid )^{-n}, \forall k,n \}\ .\]
{\bf THEOREM 1.} {\it Let $n$ is non-negative integer. Take the eigenvalue
problem for a linear differential
operator of the $k$-th order in one variable
\be
\label{e17}
 T_k \varphi \ = \ \varepsilon \varphi \ ,
\ee
where $T_k$ is symmetric. The problem (17) has $(n+1)$ linear independent
eigenfunctions in the form of polynomial in variable $x$ of the order
not higher than $n$, iff $T_k$ is quasi-exactly-solvable. The problem
(17) has an infinite sequence of polynomial eigenfunctions, iff the
operator is exactly-solvable.} \par

\noindent
{\it Comment 4.} The "if" part of the first statement is obvious.
The "only if" part is a direct corollary of Lemma 1. \par

This theorem gives a general classification of differential equations
\be
\label{e18}
 \sum_{j=0}^{k} a_j (x) \varphi^{(j)} (x) \ = \ \varepsilon \varphi(x)
\ee
having at least one polynomial solution in $x$, thus resolving the
generalized Bochner problem.
The coefficient functions $a_j (x)$ must have the form
\be
\label{e19}
a_j (x) \ = \ \sum_{i=0}^{k+j} a_{j,i} x^i
\ee
The explicit expressions (19) for coefficient function in (18) are obtained by
the substitution (2) into a general, $k$-th degree
 polynomial element of the universal
enveloping algebra $U_{sl_2({\bf R})}$.  Thus the coefficients $a_{j,i}$
can be expressed  through the coefficients of the
$k$-th degree polynomial element of the universal
enveloping algebra $U_{sl_2({\bf R})}$. The number of free parameters of the
polynomial solutions is defined by the number of parameters
characterizing a general $k$-th degree polynomial element of the universal
enveloping algebra $U_{sl_2({\bf R})}$
\footnote{ Counting free parameters, one should introduce a certain ordering
of generators to avoid double counting because of commutation relations. Also
the quadratic Casimir operator and the double-sided ideal generated by
it should not be taken into account.}. Rather straightforward calculation
leads to the following formula
\be
\label{20}
 par (T_k) = (k+1)^2
\ee
where we denoted the number of free parameters of operator $T_k$ by the symbol
$par(T_k)$.
For the case of an infinite sequence of polynomial solutions the expression
(19) simplifies to
\be
\label{e21}
a_j (x) \ = \ \sum_{i=0}^{j} a_{j,i} x^i
\ee
in agreement to the results of H.L.Krall's classification theorem
\cite{kr}(see also \cite{little}). In this case the number of free parameters
is equal to
\be
\label{22}
 par (E_k) = {(k+1)(k+2) \over 2}
\ee
In present approach Krall's theorem is simply a description of
differential operators of $k-$th order in one variable preserving
a finite flag  $ {\cal P}_0 \subset
 {\cal P}_1 \subset {\cal P}_2 \subset \dots \subset {\cal P}_k $ of spaces of
 polynomials. One can easily show that the preservation of such a set of
polynomial spaces implies the preservation of an infinite flag of such spaces.

One may ask a more general question: which non-degenerate linear differential
operators have finite-dimensional invariant sub-space of the form
\be
\label{23}
\langle \alpha (x), \alpha (x) \beta (x), \dots , \alpha (x) \beta (x) ^n
\rangle \ ,
\ee
where $\alpha (x)$ is a function and $\beta (x)$ is a diffeomorphism of the
line. Such operators are obtained from the ones of the Theorem 1
by the change of variable $x \mapsto \beta (x)$ and the "gauge"
transformation $\varphi (x) \mapsto \alpha (x) \varphi (x))$.

Let us consider the set of second order differential equations (17),
which can possesses polynomial
solutions. From Theorem 1 it follows that the corresponding differential
operator must be quasi-exactly-solvable and can be represented as
\[ T_2 =  c_{++} J^+ J^+ + c_{+0} J^+ J^0 + c_{+-} J^+ J^- + c_{0-} J^0 J^- +
c_{--} J^- J^- + \]
\be
\label{e24}
 c_+ J^+ + c_0 J^0 + c_- J^- + c ,
\ee
where $c_{\alpha \beta}, c_{\alpha}, c \in {\bf R}$.
The number of free parameters is $par (T_2) = 9$. Under the condition
$c_{++}  = c_{+0}  = c_+  =0$, the operator $T_2$ becomes
exactly-solvable (see Lemma 2) and the number of free parameters is
$par (E_2) = 6$.

{\bf LEMMA 3. } {\it If the operator (21) is such that
\be
\label{e25}
c_{++}=0 \quad and \quad c_{+} = ({n \over 2} - m)  c_{+0} \ , \ at
\ m=0,1,2,\dots
\ee
 then the operator $T_2$ preserves both ${\cal P}_n$ and
${\cal P}_m$. The number of free parameters is  $par (T_2) = 7$.} \par

{\bf PROOF.} By straightforward analysis.

In fact, the Lemma 3 means that $T_2 (J^{\alpha}(n),c_{\alpha \beta},
 c_{\alpha})$
can be rewritten as $T_2 (J^{\alpha}(m),c'_{\alpha \beta},c'_{\alpha})$.
As a consequence of Lemma 3 and Theorem 1, among polynomial solutions
of (17) there
are polynomials of order $n$ and order $m$.

{\bf Remark.}  From the Lie-algebraic point of view Lemma 3 means the existance
 of representations of second-degree polynomials in the generators (2)
possessing two invariant sub-spaces.
In general, if $n$ is not a non-negative integer in (2) ( correspondingly,
(2) becomes infinite-dimensional), then
 among representations of $k$-th degree polynomials in the generators (2),
lying in the universal enveloping algebra, there exist representations
possessing $0,1,2,...,k$
invariant sub-spaces. Also this properly implies existence
of representations of
the polynomial elements of the universal enveloping algebra,
which can be obtained starting
from different representations of the original algebra. Even starting from an
infinite-dimensional representation of the original algebra, one can construct
the elements of the universal enveloping algebra having finite-dimensional
representation (e.g. the parameter $n$ in (25) is non-integer,
however $T_2$ has the invariant sub-space of the dimension $(m+1)$). \par

Substituting (2) into (24) and then into (17), we obtain
\be
\label{e26}
P_{4}(x) \partial_x ^2 \varphi (x) \ +\ P_{3}(x) \partial_x  \varphi (x) \  +\
P_{2}(x) \varphi (x) \ =\ \varepsilon \varphi (x) ,
\ee
where the $P_{j}(x)$ are polynomials of $j$-th order with coefficients
related to  $ c_{\alpha \beta}, c_{\alpha}$ and $n$ (see (9)). In general,
problem (26) has $(n+1)$ polynomial solutions. If $n=1$, as a
consequence of Lemma 1, a  more
general spectral problem than (26) arises
\be
\label{e27}
F_{3}(x) \partial_x ^2 \varphi (x) \ +\ Q_{2}(x) \partial_x  \varphi (x) \  +\
Q_{1}(x) \varphi (x) \ =\ \varepsilon \varphi (x) ,
\ee
possessing only two polynomial solutions of the form $(ax+b)$,
where $F_3$ is an arbitrary real function of $x$ and $Q_j (x), j=1,2$ are
polynomials of order $j$. For the case $n=0$ (one polynomial solution)
the spectral problem (11) becomes
\be
\label{e28}
F_{2}(x) \partial_x ^2 \varphi (x) \ +\ F_{1}(x) \partial_x  \varphi (x) \  +\
Q_0 \varphi (x) \ =\ \varepsilon \varphi (x) ,
\ee
where $F_{2,1}(x)$ are arbitrary real functions of real $x$ and $Q_0$
is a real constant. After the transformation
\be
\label{e29}
t: \varphi \mapsto \varphi (x(z)) e ^ {A(z)} ,
\ee
where $z \mapsto x(z)$ is a diffeomorphism of the line and $A(z)$
is a certain real function, one can reduce (26)--(28) to the
Sturm--Liouville problem
\be
\label{e30}
(\partial_z ^2 \ + \ V(z) ) \varphi \ = \ \varepsilon \varphi ,
\ee
with the potential
\[ V(z) = (A')^2 + A'' + P_2 (x(z)) \]
where $A = \int ({P_3 \over P_4})dx - log z'$ for  (23). If the
functions (29), obtained after transformation, belong to the $L_2(\cal
D)$-space \footnote{In dependance on a diffeomorphism  $z \mapsto x(z)$,
the space $\cal D$ can be the infinite real line, semi-infinite real line and
a finite real interval.} ,
we reproduce the recently discovered quasi-exactly-solvable problems
\cite{t2}, where a finite number of eigenstates was found algebraically.
For example,
\be
\label{e31}
 T_2 = -4 J^0 J^- + 4a J^+ + 4b J^0 - 2(n+1) J^-
\ee
leads to the spectral problem (30) with the potential
\be
\label{e32}
 V(z) = a^2z^6 + 2abz^4 + (b^2 - (4n+3)a)z^2 ,
\ee
for which the first $(n+1)$ eigenfunctions, even in $x$
 can be found algebraically.

It is worth noting that the use of (27) as input leads to one-functional
family of the Schroedinger operators with two explicitly known eigenstates.
One of such operators has been described at \cite{jnk}, where the authors
confusingly stated about non-existance of $sl_2(\bf R)$ algebra behind.

Taking different exactly-solvable operators $E_2$ for the eigenvalue
problem (17)
one can reproduce the equations having the Hermite, Laguerre, Legendre
and Jacobi polynomials as solutions \cite{t1} \footnote{For instance, putting
the parameter $a=0$ in (31), the equation (26) converts to the Hermite
equation (after some substitution.)}. Also under special choices of general
element $E^4$, one can reproduce all
known fourth order differential equations giving rise infinite sequences of
orthogonal polynomials (see \cite{little} and other papers in this volume).

\par

  Recently, A.Gonzalez-Lopez, N.Kamran and P.Olver \cite{olver}
gave the complete description of second-order polynomial elements of
$U_{sl_2({\bf R})}$ leading to the square-integrable eigenfunctions of the
Sturm-Liouville problem (30) after transformation (29). Consequently,
for second-order ordinary
differential equations (26) the combination of this result and Theorem 1
 gives a general solution of the Bochner problem  as well as the
more general problem of classification of  equations possessing finite
number of orthogonal polynomial solutions.

{\bf 2. Finite-difference equations in one variable.}

The generalized Bochner problem is defined in the same way, as it has been
done for differential equations. The only difference is to consider
the operator $T$ in the problem (0) as a linear finite-difference
operator of a finite order. For the case of one real variable a
solution of the classification problem is very similar to the case
of ordinary differential equations described above.

Let us introduce the finite-difference analogue of the operators (2) \cite{ot}
\label{e33}
\[ \tilde  J^+ = x^2 D - \{ n \} x \]
\be
\tilde  J^0 = \  x D - \hat{n}
\ee
\[ \tilde  J^- = \ D , \]
where $\hat n \equiv {\{n\}\{n+1\}\over \{2n+2\}}$ ,
$\{n\} = {{1 - q^n}\over {1 - q}}$
is the quantum symbol, $q$ is a number characterizing the
deformation, $D z = 1 + q z D$ and $D f(z) = {{f(z) - f(qz)} \over
{(1 - q) z}} + f(qz) D$ is a shift or a
finite-difference operator (or so called the Jackson symbol (see \cite{e})).
The operators (33) after multiplication by some factors as
\[ \tilde  j^0 = {q^{-n} \over p+1} {\{2n+2\} \over \{n+1\}} \tilde J^0 \]
\[ \tilde  j^{\pm} = q^{-n/2} \tilde  J^{\pm} \]
( see \cite{ot}) form a quantum $sl_2({\bf R})_q$ algebra with the following
commutation relations
\label{34}
\[ q \tilde  j^0\tilde  j^- \ - \ \tilde  j^-\tilde  j^0 \
= \ - \tilde  j^-  \]
\be
 q^2 \tilde  j^+\tilde  j^- \ - \ \tilde  j^-\tilde  j^+ \
= \ - (q+1) \tilde  j^0
\ee
\[ \tilde  j^0\tilde  j^+ \ - \ q\tilde  j^+\tilde  j^0 \ = \  \tilde  j^+  \]
(this algebra corresponds to the second Witten's quantum deformation
of $sl_2$ in the classification of C.Zachos \cite{z}).
If $q \rightarrow 1$, the commutation relations (34) reduce
to the standard $sl_2({\bf R})$ ones. A remarkable property of generators (33)
is such that, if $n$ is a non-negative integer, they form
the finite-dimensional representation.

Similarly as for differential operators one can introduce
quasi-exactly-solvable and exactly-solvable operators.

{\bf Definition.} Let us name a linear difference operator of the $k$-th
order, $\tilde  T_k$ {\bf quasi-exactly-solvable}, if it preserves the space
${\cal P}_n$. Correspondingly, the operator $\tilde E_k$, which preserves
the infinite flag $ {\cal P}_0 \subset  {\cal P}_1 \subset {\cal P}_2
\subset \dots \subset {\cal P}_n \subset \dots $ of spaces of all
polynomials, is named {\bf exactly-solvable}.

 The analogue of the Lemma 1 holds.

{\bf LEMMA 4. } {\it (i) Suppose $n > (k-1)$.  Any quasi-exactly-solvable
operator of $k$-th order $\tilde T_k$, can be represented by a $k$-th
degree polynomial of the operators (33). If $n \leq (k-1)$, the part
of the quasi-exactly-solvable operator $\tilde T_k$ containing
derivatives up to the order $n$ can be represented by a $n$-th
degree polynomial in the generators (33).

(ii) Inversely, any polynomial in (33) is quasi-exactly solvable.

(iii) Among quasi-exactly-solvable operators
there exist exactly-solvable operators $\tilde E_k \subset \tilde T_k$.}\par

{\bf PROOF.} Straightforward analogue of the proof of the Lemma 1.

\noindent

{\it Comment 5.} If we define an analogue of the universal enveloping
algebra $U_g$ of a Lie algebra $g$ as an algebra of all polynomials
in generators $sl_2({\bf R})_q$. Then a quasi-exactly-solvable operator
$\tilde T_k$ at $k < n+1$ is simply an element of the 'universal enveloping
algebra' $U_{sl_2({\bf R})_q}$ of the algebra $sl_2({\bf R})_q$ taken
in representation (33). If $k \geq n+1$, then $\tilde T_k$ is
represented as an element of $U_{sl_2({\bf R})_q}$ plus $B D^{n+1} $,
where $B$ is any linear difference operator of the order not higher than
$(k-n-1)$. \par

Similar to $sl_2({\bf R})$ , one can introduce the grading of generators
(31) of $sl_2({\bf R})_q$ (see (15)) and, hence,of monomials of the
universal enveloping $U_{sl_2({\bf R})_q}$ (see (16)). The analogue
of Lemma 2 holds.

{\bf LEMMA 5. } {\it A quasi-exactly-solvable operator
$\tilde T_k \subset U_{sl_2({\bf R})_q}$
has no terms of positive grading, iff it is an
exactly-solvable operator.} \par

{\bf PROOF.} Straightforward analogue of the proof of the Lemma 2.

{\bf THEOREM 2.} {\it Let $n$ is non-negative integer. Take the eigenvalue
problem for a linear difference operator of the $k$-th order in
one variable
\be
\label{e35}
 \tilde T^k \varphi (x) \ = \ \varepsilon \varphi (x) ,
\ee
where $\tilde T_k$ is symmetric. The problem (35) has $(n+1)$
linear independent eigenfunctions in the form of polynomial in
variable $x$ of the order not higher than $n$, iff $T_k$ is
quasi-exactly-solvable. The problem (35) has an infinite sequence
of polynomial eigenfunctions, iff the operator is exactly-solvable
$\tilde E_k$.} \par
{\it Comment 6.} Saying the operator $\tilde T_k$ is symmetric, we imply,
that considering action of this operator on a space of polynomials of degree
not higher than $n$, one can introduce a positively-defined scalar product and
the operator $\tilde T_k$ is symmetric with respect to it . \par

{\it Comment 7.} The "if" part of the first statement is obvious. The "only if"
part is a direct corollary of Lemma 4. \par

This theorem gives a general classification of finite-difference equations
\be
\label{e36}
 \sum_{j=0}^{k} \tilde a_j (x) D^j \varphi (x) \ = \ \varepsilon \varphi(x)
\ee
having polynomial solutions in $x$.  The coefficient functions must
have the form
\be
\label{e37}
\tilde a_j (x) \ = \ \sum_{i=0}^{k+j} \tilde a_{j,i} x^i .
\ee
Particularly, this form occurs after substitution (33) into a general
$k$-th degree polynomial element of the universal
enveloping algebra $U_{sl_2({\bf R})_q}$. It guarantees the existance of at
least a finite number
of polynomial solutions. The coefficients $\tilde a_{j,i}$ are related to
the coefficients of the
$k$-th degree polynomial element of the universal
enveloping algebra $U_{sl_2({\bf R})_q}$. The number of free parameters of the
polynomial solutions is defined by the number of free parameters of a general
$k$-th order polynomial element of the universal
enveloping algebra $U_{sl_2({\bf R})_q}$\footnote{For quantum
$sl_2({\bf R})_q$ algebra
there are no polynomial Casimir operators \cite{z}. However, in the
representation (33) the relationship between generators analogous to the
quadratic Casimir operator
\[ q\tilde J^+\tilde J^- - \tilde J^0 \tilde J^0 + (\{ n+1 \}
- \hat{n}) \tilde J^0 = \hat{n} (\hat{n} - \{ n+1 \}) \]
appears. It reduces the number of independent parameters of the
second-order polynomial element of  $U_{sl_2({\bf R})_q}$. It becomes the
standard Casimir operator at $p \rightarrow 1$. }. A rather
straightforward calculation leads to the following formula
\[ par (\tilde T_k) = (k+1)^2+1 \]
(for the second order finite-difference
equation $par(\tilde T^2) = 10$). For the case
of an infinite sequence of polynomial solutions the formula (37) simplifies
\be
\label{e38}
\tilde a_j (x) \ = \ \sum_{i=0}^{j} \tilde a_{j,i} x^i
\ee
and the number of free parameters is given by
\[ par (\tilde E_k) = {(k+1)(k+2) \over 2} + 1 \]
(for $k=2$, $par(\tilde E^2) = 7$).
The increase in the number of free parameters
compared to ordinary differential equations is due to the presence of the
deformation parameter $q$. In \cite{t1} one can find the description
in present approach of the $q$-deformed Hermite, Laguerre, Legendre
and Jacobi polynomials (for definition these polynomials see \cite{e}).

The analogue of Lemma 3 holds as well

{\bf LEMMA 6. } {\it If the operator $\tilde T_2$ (see (24)) is such that
\be
\label{e39}
\tilde c_{++}=0 \quad and \quad \tilde c_{+} =( {\hat n}  - \{ m \})
\tilde c_{+0} \ , \ at \ m=0,1,2,\dots
\ee
 then the operator $\tilde T_2$ preserves both ${\cal P}_n$ and
${\cal P}_m$ and polynomial solutions in $x$ with 8 free parameters
occur.} \par

{\bf PROOF.} By straightforward analysis.

 Rather interesting situation occurs, if the parameter of deformation $q$
equals to the root of unity.

{\bf LEMMA 7. } {\it If a quasi-exactly-solvable operator $\tilde T_k$
preserves the polynomial space ${\cal P}_n$ and the parameter $q$ is satisfied
to the equation
\be
\label{e40}
q^n\ = \ 1 \ ,
\ee
then the operator  $\tilde T_k$ preserves an infinite flag
 of polynomial spaces $ {\cal P}_0 \subset  {\cal P}_n \subset {\cal P}_{2n}
\subset \dots \subset {\cal P}_{kn} \subset \dots $.}

It is worth emphasizing, that in the limit $q$ stresses to one, Lemmas 4,5,6
and Theorem 2 coincide to the Lemmas 1,2,3 and Theorem 1,respectively.
Thus the case of differential equations in one variable can be considered
as particular case of finite-difference ones. Evidently, one can consider
the finite-difference operators, which are a mixture of generators (33) with
the same value of $n$ and different $q$'s.

In closing, I would like to thank to L.Michel, V.Ovsienko, S.Tabachnikov and,
especially, V.Arnold and M.Shubin for numerous useful
discussions. Also I am very grateful to the Centre de Physique Theorique for
hospitality extended to me, where this work has been completed.

\end{document}